\documentclass{ws-ijmpe}
\usepackage[super,compress]{cite} 

\newcommand{\boldpi}{\bm{\pi}}
\newcommand{\boldtau}{\bm{\tau}}

\begin{document}

\markboth{Saori Pastore, Kuniharu Kubodera, Fred Myhrer}
           {An update of muon capture on hydrogen}

\catchline{}{}{}{}{}

\title{AN UPDATE OF MUON CAPTURE ON HYDROGEN }

\author{\footnotesize SAORI PASTORE\footnote{pastores@mailbox.sc.edu} }

\address{ Department of Physics and Astronomy,
University of South Carolina \\ 
Columbia, SC 29208, USA }

\author{FRED MYHRER\footnote{myhrer@physics.sc.edu} }

\address{Department of Physics and Astronomy,
University of South Carolina \\ 
Columbia, SC 29208, USA}

\author{KUNIHARU KUBODERA\footnote{kubodera@physics.sc.edu}}

\address{Department of Physics and Astronomy,
University of South Carolina \\ 
Columbia, SC 29208, USA }

\maketitle

\begin{history}
\received{Day Month Year}
\revised{Day Month Year}
\end{history}

\begin{abstract}

The successful precision measurement of 
%
the rate of muon capture on a proton
by the MuCap Collaboration 
allows for a stringent test of the current theoretical 
understanding of this process. 
Chiral perturbation theory, which 
is a low-energy effective field theory
that     preserves                                  
the symmetries and the pattern of 
symmetry breaking  in the underlying theory of QCD, 
offers a systematic framework for describing $\mu p$ capture
and provides a basic test of QCD at the hadronic level. 
We describe how this 
effective theory with no free parameters
reproduces the measured capture rate. 
%
A recent study has addressed new sources of uncertainties 
that were not considered in the previous works, and
we review to what extent these uncertainties are  now under control. 
Finally, the rationale for studying  
muon capture on the deuteron and some recent theoretical developments 
regarding this process are discussed. 

\end{abstract}


\ccode{PACS numbers: 12.39.Fe, 13.40.Ks, 25.30.Mr, 23.40.-s 
}


\section{Introduction} 
\label{sec:intro}

The highly precise measurement of the $\mu^-p$ capture rate provides us with 
stringent constraints on our theoretical understanding of QCD at
work
in hadrons.
The $\mu^- p$ capture occurs primarily from the hyperfine-singlet state
of  a muonic hydrogen atom~\cite{primakoff}.
%
%
The hypefine-singlet capture rate $\Gamma_0$
has 
recently been measured by the MuCap Collaboration~\cite{mucap2013}
with very high accuracy ($\sim$1 \% precision); the reported value is 
\begin{eqnarray}
\Gamma_0^{\rm exp}(\mu^- \!p \to \nu_\mu n) 
&=& 
714.9 \pm 5.4 (stat) \pm 5.1 (syst)\, {\rm s}^{-1} \, .
\label{eq:MuCapexp}
\end{eqnarray} 
Moreover, an ongoing experiment by the MuSun Group~\cite{MuSun}
envisages  to measure, with 1.5 \% precision, the $\mu^- d$ capture rate
from the hyperfine-doublet state of a $\mu-d$ atom, while the
$\mu$ -$^3$He capture rate has  been already 
measured with 0.3\% precision~\cite{exp-mu3He}. 

The recent years have witnessed a significant advancement 
in the theoretical framework of heavy-baryon chiral perturbation
theory (HB$\chi$PT), a low-energy effective field theory (EFT) of QCD.
One of the remarkable features of HB$\chi$PT
is that it allows us to systematically describe electroweak processes
involving the nucleon and light nuclei.
%
The main goal of this review article is to survey
the latest theoretical progress that has close bearing
upon the above-mentioned experimental developments
concerning muon capture on nucleons and the lightest nuclei.
This article is not intended to be a comprehensive review
of muon capture in general, and for the topics 
that are not covered here, we refer the reader
to the recent review articles of 
Refs.~\cite{FearGor04,KamKub10,Laura2012,MarKam2014}.

We give in Sec.~\ref{sec:ChPT} a highly abridged recapitulation
of HB$\chi$PT, just to provide terms and define notations needed
for this review. In Sec.~\ref{sec:psFF}, we discuss the pseudo-scalar
form factor 
that appears in the matrix element of the axial-vector current 
for the nucleon. 
The importance of radiative corrections 
along with their latest evaluations are also discussed. 
In Sec.~\ref{sec:mupcapture}, we present
the current status of theoretical calculations of the $\mu^-p$ capture rate.
Sec.~\ref{sec:two-nucleon} is devoted to a general discussion on two-nucleon
electroweak 
processes. The latest calculations of the $\mu$-$d$ capture rate
are reported in Sec.~\ref{sec:mud}, while 
discussion and a summary 
are provided in Sec.~\ref{sec:conclusion}.

\section{Heavy-Baryon Chiral Perturbation Theory}
\label{sec:ChPT}

In describing low energy-momentum hadronic phenomena
characterized by a scale $Q$                               
that is sufficiently small compared with the chiral scale 
$\Lambda_\chi\sim 1$ GeV,  
we can eliminate from the                   
Lagrangian 
those degrees of freedom that pertain to scales higher than $\Lambda_\chi$.
The resulting EFT, called chiral perturbation theory ($\chi$PT),
is a low-energy EFT of QCD~\cite{georgi}. 
The $\chi$PT Lagrangian, $\cal{L}_{\chi{\rm PT}}$, 
contains as explicit degrees of freedom 
only those hadrons that have masses
significantly lower than $\Lambda_\chi$,  
and the terms in $\cal{L}_{\chi{\rm PT}}$ are organized into 
a perturbative expansion in powers of $\epsilon=Q/\Lambda_\chi\ll 1$. 
By 
construction, $\cal{L}_{\chi{\rm PT}}$
retains all symmetries of QCD, 
including 
%
(approximate) chiral symmetry. 
The effective nature of the theory is reflected 
in the presence of low-energy constants (LECs), 
which parametrize the high-energy dynamics that
has
been eliminated (integrated out) in generating the low-energy EFT. 
If the quarks are massless, the QCD Lagrangian is chirally symmetric. 
This symmetry is spontaneously broken,
leading to the existence of massless pseudo-scalar bosons, {\it i.e.}, the
Nambu-Goldstone bosons.
In the non-strange sector of our concern here,
the Nambu-Goldstone bosons are massless pions.
Chiral symmetry is also explicitly broken by non-zero $u$ and $d$ quark masses 
which 
cause the pion to acquire a finite mass, $m_\pi$. 
Since $m_\pi \ll \Lambda_\chi$, 
the explicit chiral symmetry breaking effect
can be accounted for through an additional expansion
in the small parameter  $m_\pi/\Lambda_\chi$.  
The latter is implicit in the expansion parameter defined above, that is
$\epsilon=Q/\Lambda_\chi$,
where now Q denotes either the typical size of
the four-momentum involved in the process under consideration   
or the pion mass.

After the successful application to the meson sector~\cite{Leutwyler,Bijnens},
$\chi$PT has been extended to study processes that involve nucleons.
In the low-energy regime of interest here, it is reasonable to 
treat nucleons as non-relativistic particles, and 
accordingly we suppress
antinucleon degrees of freedom and retain only the  ``large" components 
of the nucleon field. 
The resulting theory is        
HB$\chi$PT which involves an expansion parameter 
$\epsilon^\prime=Q/m_N$ (where $m_N$ is the nucleon mass) 
in addition to the $\epsilon$ parameter defined above. 
Since $\Lambda_\chi\approx m_N$,
it is 
a
common practice to combine the expansions 
in $\epsilon$ and $\epsilon^\prime$;
thus, $n$-th order terms in HB$\chi$PT 
are those terms with a combined power of $\epsilon$ and $\epsilon^\prime$ equal to $n$. 
For review articles, we refer to,  {\it e.g.}, Refs.~\cite{bkm95,bernard08, scherer10}.

The LECs contained in the HB$\chi$PT Lagrangain
$\cal{L}_{{\rm HB}\chi{\rm PT}}$
can in principle be determined from lattice QCD calculations, 
but in practice they are fixed
by fitting appropriate experimental data.  
Once all the LECs at a given order in the expansion are determined, 
HB$\chi$PT allows us to make model-independent predictions
(to that order) on observables other than those used  
to fix the LECs. 

\section{Nucleon Pseudoscalar Form Factor}
\label{sec:psFF}

Weak processes, occurring at energies which are very small 
compared
to the weak bosons masses, can be
described 
with high accuracy
by the Fermi current-current interaction.
In particular, the weak Hamiltonian, relevant to the $\mu^- +p\rightarrow n + \nu_\mu$
reaction, is given by 
the product of the  leptonic ($L_\mu$) 
and hadronic ($J_\mu$) currents, as 
\begin{eqnarray}
H_{weak} = \frac{G_F}{\sqrt{2}} \, V_{ud} \; L_\mu J^\mu \, ,
\end{eqnarray}
where   
$G_F=1.16637(5)\times 10^{-5}$ GeV$^{-2}$ is 
the Fermi coupling constant 
while $V_{ud}=0.97418(27)$~\cite{PDG2012} is 
the CKM (Cabibbo-Kobayashi-Maskawa) matrix element. 
The leptonic current is simply 
$L_\mu = \bar{\psi}_{\nu_\mu} \gamma_\mu (1-\gamma_5) \psi_\mu$,
where $\psi_{\nu_\mu}$ ($\psi_\mu$) is the neutrino (muon) 
wave function. 
 By contrast, the hadronic current $J^\mu$ does not have a simple form
 due to complications induced by the strong interactions. 
 We can however parametrize the possible form 
 of its matrix element for a case in which 
 the initial and final states are nucleons.
 Thus, for $J_\mu = V_\mu-A_\mu$, where 
 $V_\mu$ and $A_\mu$ are the vector and axial-vector currents, respectively,
 we can write 
\begin{eqnarray}
\langle n(p^\prime ) | V_\mu | p(p)\rangle &=& \bar{u}_n(p^\prime ) 
\left[ F_V(q^2)\gamma_\mu + \frac{iF_M(q^2)}{2m_N} \, \sigma_{\mu \nu}q^\nu 
\right] u_p(p) \ ,
\\ 
\langle n(p^\prime ) |A_\mu | p(p)\rangle &=& \bar{u}_n(p^\prime )  
\left[ G_A(q^2)\gamma_\mu\gamma_5 + G_P(q^2) \frac{q_\mu }{m_\mu}\, \gamma_5 
\right]  u_p(p) \ ,
\label{eq:hadroncurr} 
\end{eqnarray} 
where $q= p^\prime-p$ is the momentum transfer
with $p$ ($p^\prime$) being the proton (neutron) momentum;
$m_N = (m_p+m_n)/2$ is the average nucleon mass,
and $m_\mu$ the muon mass.
The $F_V(q^2)$, $F_M(q^2)$, $G_A(q^2)$ and $G_P(q^2)$       
are called the vector, weak-magnetism, axial-vector 
and pseudo-scalar form factors, respectively,
and they account for the composite structure of the nucleons.
In the above expressions, we have ignored 
possible contributions from second-class currents~\cite{weinberg5}.  
The $\mu^- p$ capture reaction is the most suited  process for
obtaining information on the pseudoscalar form factor $G_P(q^2)$~\cite{primakoff}.
Bernard {\it et al.}~\cite{bkm1994} derived 
$G_P(q^2)$ 
using HB$\chi$PT at one-loop order
and obtained  
\begin{eqnarray}
G_P(q^2) &=& \frac{2m_\mu g_{\pi NN} f_\pi}{m_\pi^2-q^2} - 
\frac{1}{3} g_A m_\mu m_N \langle  r_A^2 \rangle \; ,  
\label{eq:pseudoscalar}
\end{eqnarray}
where $g_{\pi NN}$ is the strong pion-nucleon coupling constant,
and $f_\pi$ is the pion decay 
constant.
The leading term in this expression
is the well-known pion-pole term~\cite{bkm1994},
while the second  term involves the nucleon's 
mean-square isovector axial-radius, $\langle  r_A^2 \rangle $, 
which is related to the axial form factor via
$G_A(q^2) = G_A(0)\, [ 1+ \frac{1}{6} \langle  r_A^2 \rangle \, q^2 + \cdots ]$. 
%
More recently, Fearing {\it et al.}~\cite{fearing1997} also derived
Eq.~(\ref{eq:pseudoscalar})
in a slightly different HB$\chi$PT formulation.
Historically, the result 
given in 
Eq.~(\ref{eq:pseudoscalar}) was obtained
in the late sixties/early seventies by Adler and Dothan~\cite{ad66} 
using the soft-pion theorems,
and by Wolfenstein~\cite{wolfenstein} using dispersion theory.
%


A great merit of HB$\chi$PT is that it allows
us
to estimate the size of errors
associated with a given theoretical calculation. In the case of the nucleonic
pseudoscalar form factor, corrections at two-loop order
have been explicitly evaluated by Kaiser~\cite{kaiser03}, and 
found to
be negligible, provided that the involved LECs were of natural size. 
%
When we insert in Eq.~(\ref{eq:pseudoscalar}) the momentum transfer 
pertaining to the $\mu^- p$ capture reaction,  {\it i.e.}
$q^2 = -0.88 m_\mu^2$, along with the experimentally determined axial radius~\cite{raexp1999} 
$\langle  r_A^2 \rangle = 0.44 \pm 0.02$ fm$^2$, 
HB$\chi$PT at one-loop order gives $G_P(q^2=-0.88m_\pi^2) = 8.26 \pm 0.23$,
which is in excellent agreement with the empirical value of
$8.06 \pm 0.55$, obtained by 
the recent MuCap experiment~\cite{mucap2013}. 
The details of the framework used in obtaining 
this experimental value was thoroughly reviewed 
in Ref.~\cite{KamKub10}  
It should be stressed
that, in order to
match the 1\% accuracy achieved in the measurement of the $\mu^- p$ capture rate,
radiative corrections need to be carefully taken into account;
the MuCap group used the radiative corrections 
evaluated by Czarnecki {\it et al.}~\cite{czarnecki2007}.             
Since the time when
Ref.~\cite{KamKub10}  was written,    
there have been  significant developments which affect the theoretical description
of the $\mu^- p$ capture reaction, 
and these developments are reviewed in the next section.

\section{The $\mu^- p$ Capture Rate}
\label{sec:mupcapture}

The 1\% experimental accuracy 
achieved by the MuCap Collaboration~\cite{mucap2013} 
in the measurement of $\Gamma_0$, 
poses a challenge for the theory.
To attain a comparable theoretical precision,
higher-order HB$\chi$PT contributions,
including radiative corrections, need to be
accounted for.
In HB$\chi$PT the $\mu^- p$ capture rate has been evaluated by 
Fearing {\it et al.}~\cite{fearing1997}, Ando {\it et al.}~\cite{ando2000}, 
and Bernard {\it et al.}~\cite{bhm2001}. 
In these works the transition amplitude was evaluated 
including $m_N^{-1}$ nucleon recoil corrections entering
at next-to-leading order (NLO).   
At next-to-next-to-leading order (N$^2$LO), there are 
recoil  corrections  of order $m_N^{-2}$ 
as well as loop corrections. 
Since all the LECs at N$^2$LO are known, 
HB$\chi$PT leads to model-independent predictions for 
the $\mu^- p$ capture rate. 
Based on the convergence pattern exhibited by the contributions to the capture rate 
evaluated in, {\it e.g.}, Ref.~\cite{bhm2001}, 
it is estimated that N$^3$LO corrections would
contribute at the 1\% level to the capture rate. 
Comparison of the results for $\Gamma_0$ obtained in HB$\chi$PT
with the earlier results obtained 
in the phenomenological approach, {\it e.g.}, Refs.~\cite{primakoff,opat}
can be found in Refs.~\cite{FearGor04,KamKub10}.

A recent HB$\chi$PT calculation
of $\mu^- p$ capture~\cite{udit2013} takes into account
radiative corrections of order $\alpha\sim 1/137$,
which enter at N$^2$LO in the chiral expansion, 
that is, they scale as $\left(Q/\Lambda_\chi\right)^2$;
%
the fact that $Q \sim m_\mu$ in $\mu^-p$ capture
leads to the relation 
$(Q/\Lambda_\chi)^2\sim (m_\mu/\Lambda_\chi)^2\sim1/100\sim\alpha$. 
%
These radiative corrections include 
standard QED vacuum polarization effects~\cite{EirasSoto}, 
electroweak loop corrections, as well as proton 
finite-size corrections~\cite{Friar}. 
Divergences generated by electroweak loops appearing at N$^2$LO are 
regulated by electroweak LECs,  
which describe  short-distance effects. 
These LECs    represent  the 
``inner" corrections in the formalism of  Sirlin~\cite{sirlin1967}, and  
%
%
are determined by matching the 
expressions for the neutron $\beta$-decay radiative corrections
obtained by Marciano and Sirlin~\cite{MarSir86},
and those derived in HB$\chi$PT by Ando {\it et al.}~\cite{ando2004}. 
%
%
The radiative corrections derived in Ref.~\cite{udit2013} are found to 
be in agreement with those evaluated by Czarnecki {\it et al.}~\cite{czarnecki2007},
which have been used by the MuCap Collaboration. 
In Ref.~\cite{udit2013}, it was also found that
electroweak loop-corrections increase the calculated rate 
$\Gamma_0$ by as much as $\sim$ 2\%, 
an increase that, due to partial cancellations 
among other terms~\cite{udit2013}, is dominated by the aforementioned electroweak LECs. 
%
In addition, Raha {\it et al.}~\cite{udit2013} showed that,
even if we generously assign a 10\% uncertainty 
to the nucleon isovector axial radius, 
$\langle r_A^2\rangle^{1/2}$, 
the corresponding variation in $\Gamma_0$ is within $\sim$0.5\%.

Apart from the above-mentioned $\sim$1 \% uncertainty 
due to N$^3$LO  contributions,
the N$^2$LO calculation of $\Gamma_0$
involves additional uncertainties. 
These arise from uncertainties associated with 
the nucleon axial-vector coupling constant, $g_A$,
and the nucleon-pion coupling constant, $g_{\pi NN}$. 
The axial constant 
$g_A$ is determined most directly 
from the measured asymmetry parameter $A$ 
in neutron beta decay~\cite{Mund2013,Mendenhall2013}. 
Historically, the value of $g_A$ 
recommended by the Particle Data Group (PDG) 
has steadily increased, and the latest reported value is 
$g_A=1.2701 \pm 0.0025$~\cite{PDG2012}.  
Unfortunately,  this is not the last word in the saga of $g_A$. 
The asymmetry parameter $A$ in neutron beta decay 
has recently been 
re-measured
by two groups~\cite{Mund2013,Mendenhall2013},
and they have obtained $g_A\simeq 1.276$, 
which is larger than the PDG2012 value~\cite{PDG2012}. 
It should be noted that  the value                          
$g_A\simeq 1.276$ is more consistent with the     
smaller value of the neutron life time, $\tau_n= 880.0\pm 0.9$ s,
which is now recommended by the PDG~\cite{PDG2012};
%
see the arguments in Ref.~\cite{greene2011}
advocating for a smaller value of $\tau_n$. 
The relation between the new larger value of  $g_A$ 
and the smaller $\tau_n$  
has also been discussed in  
Refs.~\cite{Mund2013,ivanov2013}. 
We note  that the value of the neutron lifetime  is 
not settled experimentally, as 
shown by Yue {\it et al.}~\cite{yue}. 
The  pion-nucleon coupling constant $g_{\pi NN}$ 
has been extracted from both 
nucleon-nucleon 
and pion-nucleon scattering data,
as discussed recently in,  
{\it e.g.}, Refs.~\cite{Ericson2002,Arndt2004,Bugg2003,Arndt2006,Baru2011,Baru2}. 
No consensus has been reached on the 
value of $g_{\pi NN}$, 
and the best we can do at present 
is to allow $g_{\pi NN}$ to have a range,
$g_{\pi NN} = 13.044$---$13.40$;
the smaller value is taken from Ref.~\cite{Nijmegen1997}
and the larger one from Ref.~\cite{KH83}. 
The uncertainty in $g_{\pi NN}$ 
affects the evaluation of $\Gamma_0$ at N$^2$LO
via the Goldberger-Treiman discrepancy, 
$\Delta_{GT} = g_A \, m_N /(g_{\pi NN} f_\pi ) -1$.

Given the changing value of  $g_A$
and the existing uncertainty in $g_{\pi NN}$, 
it is important to estimate variations in $\Gamma_0$
due to changes in $g_A$ and $g_{\pi NN}$.
Such an estimation has been carried out by 
Pastore {\it et al.}~\cite{saori},
and their results are shown in Table~\ref{tab:predictions}. 
Note that 
all the theoretical values for $\Gamma_0$ in Table~\ref{tab:predictions} 
are within the experimental errors given in Eq.~(\ref{eq:MuCapexp}). 
If we use the latest published values for 
$g_A$ and $g_{\pi NN}$~\cite{Mund2013,Baru2011,Baru2}, 
the larger $\Gamma_0$ in the last row appears 
theoretically favored.  
Variations in the calculated value of $\Gamma_0$
due to the existing uncertainties in $g_A$ and $g_{\pi NN}$
are of comparable size to 
the estimated contributions from N$^3$LO terms~\cite{ando2000,bhm2001,saori}. 
Therefore, it does not seem warranted at present
to go on to N$^3$LO calculations, 
which involve a major effort. 

\begin{table}
\begin{center}
\caption{Variations of the $\mu^- p$ capture rate $\Gamma_0$ in s$^{-1}$ 
and the Goldberger-Treiman discrepancy, $\Delta_{GT}$, are given 
with respect to 
some selected values for 
$g_A$ and $g_{\pi NN}$. 
The radiative corrections discussed in 
Ref.~\cite{udit2013} are accounted for. 
}
\label{tab:predictions}
\begin{tabular}{c|c||c||c}
\hline\hline
$g_A$   &    $g_{\pi NN}$ &   $\Delta_{GT}$ 
&\,$\Gamma_0 $ 
\\[1ex]
\hline
1.266    &     13.40   & -0.040 
& 707.1  \\ 
1.2761   &     13.40   & -0.036   
& 715.8  \\ 
\hline
1.266    &     13.044   & -0.014 
& 710.4 \\ 
1.2761   &     13.044   & -0.006 
& 719.2 \\ 
\hline
\hline
\end{tabular}
\end{center}
\end{table}
%

\section{Family of Two-Nucleon Weak-Interaction Processes}
\label{sec:two-nucleon}

There exists a long list of literature on the evaluation
of the  $\mu^- d$ capture rate~\cite{Ando2000,pisa2011,pisaPRL,Adam};
the most recent works are  strongly motivated 
by the ongoing experimental effort 
by the MuSun collaboration~\cite{MuSun} at PSI,
which aims at measuring it at 1.5\% precision. 
In the recent theoretical developments,
HB$\chi$PT has been playing an important role, 
as 
described below.

The extension of HB$\chi$PT to multi-nucleon systems
is accomplished following the scheme 
formulated 
by Weinberg in Refs.~\cite{weinberg,weinberg2,weinberg3,weinberg4}.
%
%
The basic idea is to categorize Feynman diagrams describing 
a given reaction into irreducible and reducible diagrams. 
Irreducible diagrams are those that do not
involve pure nucleonic intermediate states,
and all other diagrams are called reducible.
Let us consider a two-nucleon system as an example.
The HB$\chi$PT nucleon-nucleon potential, $v_{ij}^{\rm EFT}$,
is defined as the sum of all the irreducible diagrams 
entering the $NN\rightarrow NN$ transitions amplitude. 
The contributions of reducible diagrams can be  included 
by solving the Schr\"{o}dinger equation 
in which $v_{ij}^{\rm EFT}$ appears as the potential.
The HB$\chi$PT three-nucleon potential, $v_{ijk}^{\rm EFT}$,
can be defined in a similar manner.
For an $A$-body system, the nuclear wave function
$\Phi^{\rm EFT}$ is a solution of the $A$-body
Schr\"{o}dinger equation with the Hamiltonian given by 
\begin{eqnarray}
H^{\rm EFT}=\sum_{i=1}^AK_i
+\sum_{i<j} ^Av_{ij}^{\rm EFT}
+\sum_{i<j<k}^Av_{ijk}^{\rm EFT}+\dots \; , 
\end{eqnarray}
where $K_i$ is the kinetic energy of the $i$th nucleon;
the dots denote operators involving more than three nucleons,
which are of higher order in the HB$\chi$PT expansion
and hence                                              
can be dropped. 
%
The matrix element
of a nuclear electroweak transition 
is given by
\begin{equation}
{\cal M}^{\rm EFT}= \langle\Phi_{f}^{\rm EFT}|
\sum_i^A{\cal O}_i^{\rm EFT}
+\sum_{i<j}^A{\cal O}_{ij}^{\rm EFT}+\dots|
\Phi_i^{\rm EFT}\rangle\,,\label{eq:MEFT}
\end{equation}
where the initial and final wave functions are obtained 
in the manner described above.
The transition operators can have terms involving three or more nucleons,
but they are of higher orders in the HB$\chi$PT expansion. 
The one-body (two-body) transition operator,
 ${\cal O}_i^{\rm EFT}$ (${\cal O}_{ij}^{\rm EFT}$),   
is obtained as the sum of all irreducible diagrams involving 
the relevant external current for one-nucleon (two-nucleon) diagrams. 
The derivation of these operators in HB$\chi$PT 
was pioneered by Park, Min, and Rho 
in Ref.~\cite{ParkMinRhoEM} for the electromagnetic current,
and in Ref.~\cite{ParkMinRhoAX} for the weak axial current; 
%
${\cal O}_i^{\rm EFT}$ and ${\cal O}_{ij}^{\rm EFT}$       
were derived up to N$^3$LO. 
At this order the two-body 
operators ${\cal O}_{ij}^{\rm EFT}$ 
include contributions from one- and two-pion exchanges.      
More recently, chiral electromagnetic current (and charge) operators at one-loop order
have been derived by K\"{o}lling {\it at al.} 
in the unitary transformation method~\cite{Kolling09,Kolling11},
and by Pastore {\it et al.} 
within time-ordered perturbation theory~\cite{Pastore08,Pastore09,Piarulli12}. 
These two approaches differ among themselves and 
from the scheme adopted by Park {\it et al.},
in the treatment of the reducible contributions. A discussion on these differences
can be found in Refs.~\cite{Kolling11,Pastore08,Pastore09,Piarulli12}. A derivation of the axial 
current within the formalism developed in Refs.~\cite{Pastore08,Pastore09,Piarulli12} is being 
vigorously pursued~\cite{Baroni14}.
%

\begin{figure*}[t]
\begin{center}
\includegraphics[height=2.50cm,keepaspectratio]{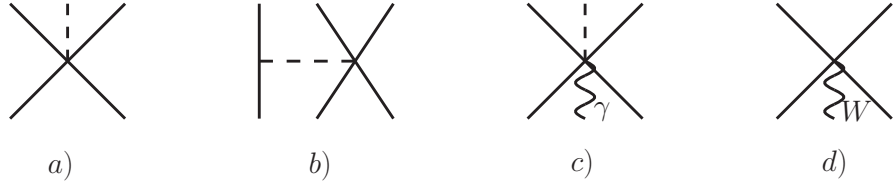}
\caption{
\label{ct}  Diagrams involving the  LEC, $d^R$. 
The solid and dashed lines represent nucleons and pions, respectively.
      The wavy line in diagram c) [diagram d)] represents a photon ($W$ weak boson).}
\end{center}
\end{figure*}

In considering the specific case of $\mu^- d$ capture,
we note the following two crucial points:
(i) for the low-energy Gamow-Teller (GT) transition 
which governs this process, 
the  one-body transition 
operator, ${\cal O}_i^{\rm EFT}$,
is well known, see Eq.~(\ref{eq:hadroncurr}); 
(ii) the two-body terms, ${\cal O}_{ij}^{\rm EFT}$,
involve only one unknown LEC, 
which in the literature is denoted by $d^R$. 
This LEC parameterizes the strength 
of a contact-type four-nucleon coupling
to the axial current;
diagram d) in Fig.~\ref{ct} illustrates this coupling. 
Thus $d^R$ can be regarded as the 
two-nucleon analog of the nucleon axial-vector 
coupling constant $g_A$. 

As noted by Park {\it et al.}~\cite{park2}, $d^R$ also enters the two-body GT amplitude 
of the solar $pp$ fusion reaction, tritium $\beta$-decay~\cite{park2,marcucci2013}, 
and $\nu d$ scattering at low energies~\cite{Nakamura:2002jg}. 
This means that, if $d^R$ 
can be determined 
%
from the experimentally known rate                 
of any one of these processes, robust predictions
can be made for the remaining reactions.
Moreover, $d^R$ enters pure hadronic as well as electromagnetic 
reactions. To the former class belong, for example, the 
processes
represented by
diagrams a) and b) in Fig.~\ref{ct}. 
Diagram~a) appears in 
the hadronic reaction $NN\to NN\pi$~\cite{HvKM2000,review}.
Diagram~b) contributes to three-nucleon interactions,
giving rise to a relation between $d^R$ and $c_D$,
an LEC that parameterizes the short-range contribution 
to the three-nucleon 
potential~\cite{pdchiral,Epelbaum:2008ga,Machleidt:2011zz,gazit}.
%
Diagram c) represents an electromagnetic process
that involves $d^R$.
This diagram appears in,  
{\it e.g.}, the $\gamma d\!\to\!\pi NN$ reaction~\cite{gammad,gammad_nn} 
and  $\pi^- d\!\to\! \gamma NN$  reaction~\cite{gardestig,phillips}.
The last reaction has long been known 
as a tool to extract  the $nn$-scattering length,
and a detailed HB$\chi$PT study of this extraction procedure
has recently been made 
by Gardestig and Phillips~\cite{Gardestig:2005pp,gardestigreview}. 
 
In all the reactions given above,
the short-ranged operator accompanied by the LEC $d^R$
parameterizes common short-distance two-nucleon physics
that has been integrated out. 
%
%
How these processes are interconnected 
can be easily understood by examining
the structure of the chiral Lagrangian,
which is customarily written in terms of 
the  chiral field $U(x)$.\footnote{
Ref.~\cite{ParkMinRhoAX} defines $U(x)={\rm exp}[i\boldtau \! \cdot\! \boldpi (x)/f_\pi]$, 
whereas Ref.~\cite{bkm95}  uses the ``sigma-gauge" expression 
$U(x)=\sqrt{1-(\boldpi/f_\pi)^2} +i\boldtau \! \cdot \!\boldpi(x)/f_\pi$. 
}
%
The contact interaction, illustrated by the diagrams of Fig.~\ref{ct}, 
is given by a four-nucleon interaction Lagrangian of the form 
\begin{eqnarray}
{\cal L}_{NN}= -2 d\; \left(N^\dagger S\!\cdot\! u \,N\right) 
N^\dagger\!N      \, , 
\label{lag}
\end{eqnarray} 
where $N(x)$ is the heavy-nucleon field, $S^\mu$ is the nucleon 
covariant spin operator, and     
$ u_\mu\equiv i \,  (\xi^\dagger \partial_\mu 
\xi- \xi \partial_\mu \xi^\dagger)$, 
with
$\xi = \sqrt{U(x)}$.  
The coupling constant $d$ becomes $d^R$
after the renormalization procedure is implemented.
Including the external electroweak currents,  $V_\mu$ and $A_\mu$,
we can see
that $u^\mu$ connects the 
pion emission vertex       
with the external vector and axial-vector currents via 
\begin{eqnarray}
f_{\pi} { u_{\mu}}= - {\boldtau} 
\partial_{\mu} {\boldpi} 
- \varepsilon_{3ab}{ V_{\mu}} \pi_a \tau_b +f_{\pi}{ A_{\mu}}+ \cdots \ ,
\label{u}
\end{eqnarray} 
where  
the ellipses represent higher powers in the pion field. 
The contributions of the first term in Eq.~(\ref{u}) 
to the contact Lagrangian, ${\cal L}_{NN}$, 
give rise to the vertices appearing in diagrams 
a) and b) in Fig.~\ref{ct},
while the second and third terms in Eq.~(\ref{u})
generate the vertices appearing in diagrams c) and d).
%
%
%

The first determination of the LEC $d^R$ from experimental data 
was done in Ref.~\cite{park2} by reproducing the tritium $\beta$-decay rate,
$\Gamma_\beta^t$.
 %
%
 In a recent work Gazit {\it et al.}~\cite{gazit} used 
 the $^3$H and $^3$He binding energies
 as well as $\Gamma_\beta^t$ to fix  $d^R$.
 Although there are good reasons to believe that the 
 determination of $d^R$ with the use of observables 
 in the $A=3$ systems is reliable to the quoted level,  
 it is desirable to determine $d^R$
 within the two-nucleon system 
 without resorting to the 
 input from the three-nucleon system.     
 As discussed in the next section,
 the high-precision measurement
 of the $\mu^- d$ capture rate by the MuSun group~\cite{MuSun} 
 is of particular importance    for              
 the determination of $d^R$.    

\section{Muon-Deuteron Capture Rate}
\label{sec:mud}

Recent experimental 
and theoretical developments have reached such a level of accuracy that
all the relevant LECs are controlled with reasonable precision.
Meanwhile, HB$\chi$PT studies of the two-nucleon systems
have established that the low-energy weak-interaction processes
in the $A$=2 systems involve only one unknown LEC, $d^R$,  
up to N$^3$LO~\cite{ParkMinRhoAX}.
This means that if we can carry out an explicit calculation of
${\cal M}^{\rm EFT}$ in Eq.(\ref{eq:MEFT}),
and if sufficiently accurate data on $\mu^- d$ capture becomes available, 
then $d^R$ can be fixed.                          
This will allow us to correlate
in a reliable model-independent manner, 
all the low-energy electroweak processes in the two-nucleon systems.
The on-going high-precision measurement of the $\mu^- d$-capture rate
by the MuSun Collaboration is expected to play 
an important role in this program;
cf. {\it e.g.}, Refs.~\cite{Ando2000,MuSun,pisa2011,pisaPRL,Adam}.  

To set the stage for surveying the recent developments 
based on HB$\chi$PT, 
we first briefly describe the traditional method
known as the standard nuclear physics approach (SNPA).
SNPA starts with the assumption that an $A$-nucleon system
is described by the Hamiltonian 
\begin{eqnarray}
H^{\rm SNPA} =\sum_{i=1}^AK_i
+\sum_{i<j} ^Av_{ij}^{\rm SNPA}
+\sum_{i<j<k}^Av_{ijk}^{\rm SNPA} \; , 
\label{eq:SNPA}
\end{eqnarray}
%
%
where $v_{ij}^{\rm SNPA}$ ($v_{ijk}^{\rm SNPA}$) is a 
high precision phenomenological 
two-body (three-body) potential. 
These potentials are constrained by reproducing 
existing two-nucleon scattering data as well as 
the binding energies and level structure of light nuclei, etc.;
see, {\it e.g.,} Refs.~\cite{Argonne,3N}.
The electroweak transition operators in SNPA consist of 
one-body impulse-approximation (IA) terms,
and two-body meson exchange-current (MEC) terms;
the IA terms can be determined from the coupling of a single nucleon 
to the electroweak current, while the MEC terms are derived 
from boson-exchange diagrams.  
SNPA has been applied with great success to the description
of nuclear observables in light nuclei, see, {\it e.g.}, Ref.~\cite{Rocco}. 
Detailed calculations of $\mu^- d$ capture based on SNPA 
were carried out by Tatara {\it et al.}~\cite{Tatara} 
and by Adam {\it et al.}~\cite{Adam-Q} 
more than twenty years ago.
Tatara {\it et al.} obtained for the hyperfine-doublet capture rate
$\Gamma_d({\rm SNPA}) =300 - 400$ s$^{-1}$,
and it was noted that more than 50 \% 
of the contributions to $\Gamma_d$
come from higher partial-wave states 
($L\ge 1$) for the final two-nucleon relative motion.
Even though SNPA is believed to work with reasonable accuracy,
it involves a certain degree of model dependence. 
In principle, HB$\chi$PT  should allow us to treat multi-nucleon 
systems in a model-independent way.

In the past it was a challenge to generate, 
strictly within the EFT framework,
nuclear wave functions with accuracy 
comparable to that of the SNPA nuclear wave functions.
To avoid this difficulty,
Park {\it et al.}~\cite{park2}
proposed to replace $\Phi^{\rm EFT}$ in Eq.(\ref{eq:MEFT})
with $\Phi^{\rm SNPA}$,
where $\Phi^{\rm SNPA}$ is 
a phenomenological nuclear wave function
obtained as an exact eigenstate of the                   
nuclear Hamiltonian $H^{\rm SNPA}$ in Eq.~(\ref{eq:SNPA}).
This hybrid method, termed EFT$^*\!$, has the advantage
that it can be applied to complex nuclei ($A=3,4, \ldots$)
with essentially the same precision 
as to the $A=2$ case;
it thus allows us to fix $d^R$ from observables 
pertaining to complex nuclei as was done in, {\it e.g.}, Ref.~\cite{park2}.

To achieve a theoretical accuracy compatible with the expected precision 
of the MuSun experiment one must evaluate the $\mu^- d$ capture rate 
in HB$\chi$PT at least to N$^2$LO~\cite{Ando2000,pisa2011,pisaPRL,Adam}.
An EFT$^*$-based calculation of $\mu^- d$ capture was 
carried out by Ando {\it et al.}~\cite{Ando2000}, who used the value 
of $d^R$ 
%
obtained in Ref.~\cite{park2}
by applying EFT$^*$ to tritium beta decay;
Ando {\it et al.} report the value $\Gamma_d({\rm EFT}^*\!)=386\,{\rm s}^{-1}$.
%
We remark en passant that,
in deriving the so-called fixed terms of orders $m_N^{-1}$ and $m_N^{-2}$,
Ando {\it et al.}~\cite{Ando2000} used the
Foldy-Wouthuysen transformation 
instead of the non-relativistic heavy baryon expansion.
The two methods are not identical  
but  one scheme can be transformed to the other as shown in, 
{\it e.g.,} Ref.~\cite{gardestig07}. 
To order $m_N^{-2}$, the results of the two methods are identical.  
Note that in this EFT$^*$ calculation of $\Gamma_d$ 
the weak transition operators are derived in HB$\chi$PT 
whereas  the two-nucleon wave functions are obtained using 
the Argonne $v18$ potential~\cite{Argonne}. 
The high-momentum components of this $NN$ potential is          
regulated 
by a Gaussian cut-off function. 
The inclusion of such regularization can in principle 
cause the violation of CVC (the conservation of the vector current) 
and PCAC (partial conservation of the axial current).
Furthermore, the value of the LEC, $d^R$, becomes dependent
on this regularization procedure, a topic which is  
also discussed in Ref.~\cite{Truhlik2010}. 
However, if the numerical results for the observable $\Gamma_d$
turns out to be stable 
against changes in the cutoff parameter,
it is reasonable to conclude that,
despite the above-mentioned formal issues,
an EFT$^*$ calculation of $\Gamma_d$ is practically model-independent.

The most detailed study to date 
of  the $\mu^- d$  capture rate was made by 
Marcucci {\it et al.}~\cite{pisaPRL,pisa2011},
who carried out calculations based 
on both SNPA and HB$\chi$PT.                      
Their work also includes the calculation of 
the $\mu ^3$He capture reaction.
In their SNPA calculation, Marcucci {\it et al.} used 
the initial and final nuclear wave functions 
for the $A=2$ and 3 derived from the  
Argonne $v18$ two-nucleon potential~\cite{Argonne}
(in combination with the Urbana IX three-nucleon potential~\cite{3N} 
in the case of $A=3$). 
The relevant weak-interaction transition operators were obtained
using SNPA~\cite{pisaPRL,pisa2011}, which involves one parameter,
the $N$-$\Delta$ axial coupling constant that controls 
the two-body axial exchange current.
After fixing this coupling constant by reproducing $\Gamma_\beta^t$, 
Marcucci {\it et al.} obtained 
$\Gamma_d({\rm SNPA})=390.4 \sim 390.9$ s$^{-1}$, 
the lower (higher) value
corresponding to the use of $g_A=1.2654$ ($g_A=1.2695$).
It is to be noted that, the dependence of the results
on the adopted value of $g_A$ is significantly reduced 
because of the constraint that the experimental value
of $\Gamma_\beta^t$ be reproduced for each choice of $g_A$.

In their HB$\chi$PT calculation,
Marcucci {\it et al.}~\cite{pisa2011,pisaPRL} 
used nuclear wave functions generated          
by the chiral N$^3$LO two-nucleon potential~\cite{Machleidt:2011zz}, 
supplemented with the chiral N$^2$LO three-nucleon potential~\cite{gazit}
in the case of $A=3$.
The transition operators were derived to N$^3$LO,
which included two-pion exchange currents. 
To this order the theory still contains only one unknown LEC,
$d^R$.  This LEC was determined by reproducing 
$\Gamma_\beta^t$.
In the spirit of low-energy EFT, 
Fourier transformation from momentum- to coordinate-space 
was regulated with a Gaussian regulator with a cutoff $\Lambda$,
which was taken to be $\Lambda=500-800$ MeV, 
following 
Park {\it et al.}~\cite{park2}.
As mentioned, the stability of the results against the change of $\Lambda$
is considered to give a measure of model-independence.
Marcucci {\it et al.}~\cite{pisa2011,pisaPRL} obtained 
$\Gamma_d({\rm EFT}) =393.6(7)$ s$^{-1}$ 
with practically no $\Lambda$-dependence within the 
range $\Lambda=500-800$ MeV.
Combining the results of their SNPA and HB$\chi$PT calculations,
Marcucci {\it et al.} concluded that the model dependence
due to interactions, currents, and the cutoff $\Lambda$ 
is at the 1 \% level, and they gave as the best estimate the value
$\Gamma_d=(389.7-394.3)$ s$^{-1}$.

At this order, like in the case of $\mu^- p$ capture, 
the radiative corrections need to be carefully studied. 
The HB$\chi$PT-based evaluation of radiative corrections 
for $\mu^- d$ capture is yet to be completed~\cite{Song2013}.

\section{Discussion and Summary}
\label{sec:conclusion}

A topic closely related to muon capture on  hydrogen
is that of muon capture on $^3$He.
A                                 
measurement of this capture rate
gave $\Gamma(\mu ^3{\rm He}) = 1496$ s$^{-1}$ 
with 0.3\% precision~\cite{exp-mu3He}.  
An EFT$^*$ calculation of $\mu ^3$He capture 
was carried out by Gazit {\it et al.}~\cite{gazit1,gazit}, 
who used the Argonne $v18$ NN interactions~\cite{Argonne} 
and the Urbana IX three-nucleon potential~\cite{3N}.
Most recently, Marcucci {\it et al.}~\cite{pisa2011,pisaPRL}  evaluated 
$\Gamma(\mu^3{\rm He})$ 
in both SNPA and HB$\chi$PT,
and they
found good agreement 
between the SNPA and HB$\chi$PT results,
similarly to the case of  $\mu d$ capture.
Marcucci {\it et al.} reported $\Gamma(\mu^3{\rm He}) = 1494 \pm 21$ s$^{-1}$.
Radiative corrections obtained 
in the Marciano-Sirlin method~\cite{czarnecki2007}  
were used in arriving at this value.
Agreement between theory and experiment is very satisfactory. 

The $\mu^- p$ capture reaction,
$\mu^-\!+\!p\!\to\!\nu_\mu\!+\!n$, discussed in Sec.~\ref{sec:mupcapture} 
is often called 
ordinary muon capture (OMC) in contradistinction
to radiative muon capture (RMC),
$\mu^-\!+\!p\!\to\!\nu_\mu \!+\!n\!+\!\gamma$.
It is noteworthy that the study of RMC
in principle allows the determination of
the $q^2$ dependence of $G_P(q^2)$ 
appearing in Eq.(\ref{eq:pseudoscalar}). 
For an obvious reason, however, 
RMC has a much smaller branching ratio
than OMC, and for a longtime it was a great experimental challenge
to observe RMC.
Wright {\it et al.}~\cite{RMCexp} succeeded in measuring
the highly suppressed RMC rate. 
However, the  $G_P(q^2)$ extracted by Wright {\it et al.}
is larger than what was derived from other experiments~\cite{KamKub10,FearGor04}. 
Furthermore, the measured RMC capture rate  
does not agree with the theoretical value 
obtained in HB$\chi$PT~\cite{meissner1998,ando1998,bhm2001}. 
This RMC experimental result                                 
remains a puzzle;  
see the discussions in the reviews~\cite{KamKub10,FearGor04} 
for more details. 

The high-precision measurement  of 
the hyperfine-singlet $\mu^- p$ capture rate $\Gamma_0$
by the MuSun Group has been conducive to intensive theoretical studies
of this reaction based on HB$\chi$PT.
Recent developments include the HB$\chi$PT calculation 
of the radiative corrections by Raha {\it et al.}~\cite{udit2013}, and
Pastore {\it et al.}'s work~\cite{saori} 
on the propagation of uncertainties in the empirical values of 
the coupling constants, $g_A$ and $g_{\pi NN}$
to uncertainties in the calculated value of 
$\Gamma_0$. 
Pastore {\it et al.}~\cite{saori}
report
$\Gamma_0= 718 \pm 7$ s$^{-1}$, which is in good agreement with the 
experimental value given in Eq.(\ref{eq:MuCapexp}). 

As for $\mu^- d$ capture, 
Marcucci {\it et al.}'s  HB$\chi$PT calculation of $\Gamma_d$
is reported to have 1\% accuracy, which matches 
the experimental accuracy  of 1.5 \% expected  
in the on-going MuSun measurements.        
Marcucci {\it et al.}~\cite{pisa2011,pisaPRL}  
used  the radiative corrections 
calculated in the Sirlin-Marciano approach~\cite{czarnecki2007}. 
%
It is desirable to derive 
these radiative corrections  within the HB$\chi$PT framework. 
Such a calculation is currently underway~\cite{Song2013}.

\section*{Acknowledgements}  

This work is supported in part by the National Science Foundation, 
Grant No. PHY-1068305.  

\appendix  






\begin{thebibliography}{99} 

\bibitem{primakoff} 
H. Primakoff, in {\it Nuclear and Particle Physics at Intermediate Energies} 
ed. J.B. Warren 
(Plenum, New York, 1975) p.1. 

\bibitem{mucap2013} 
V. A. Andreev {\it et al.} (MuCap Collaboration), 
{\it Phys. Rev. Lett.} {\bf 110} (2013) 012504. 

\bibitem{MuSun}
  V.~A.~Andreev {\it et al.}  [MuSun Collaboration],
  arXiv:1004.1754 [nucl-ex].

\bibitem{exp-mu3He}
R. Ackerbauer {\it et al.}, {\it Phys. Lett B} {\bf 417} (1998) 224. 

\bibitem{KamKub10}
P. Kammel and K. Kubodera {\it Annu. Rev. Nucl. Part. Sci.} {\bf 60} (2010) 327.

\bibitem{FearGor04}
T. Gorringe and H. W. Fearing  {\it Rev. Mod. Phys.} {\bf 76} (2004) 31. 

\bibitem{Laura2012}
L.E. Marcucci, {\it Int. J. Mod. Phys. A} {\bf 27} (2012) 1230006  
[arXiv:1112.0113].

\bibitem{MarKam2014}
P. Kammel and L. E. Marcucci, in preparation. 

\bibitem{georgi}
H. Georgi, {\it Weak Interactions and Modern Particle Theory},
Addison-Wesley Publ. Comp. (NY, 1984), p. 80. 

\bibitem{Leutwyler}
J. Gasser and H. Leutwyler, {\it Phys. Rep.} {\bf 87} (1982) 77; 
{\it Ann. Phys.} {\bf 158} (1984) 142. 

\bibitem{Bijnens}
J. Bijnens, {\it Prog. Part. Nucl. Phys.} {\bf 58} (2007) 521.

\bibitem{bkm95} 
V. Bernard, N. Kaiser and U.-G. Meissner,  
{\it Int. J. Mod. Phys. E} {\bf 4} (1995) 193.

\bibitem{bernard08}
V. Bernard, {\it Prog. Nucl. Part. Phys.} {\bf 60} (2008) 82.

\bibitem{scherer10}
S. Scherer, {\it Prog. Nucl. Part. Phys.} {\bf 64} (2010) 1. 

\bibitem{PDG2012}  
J. Beringer {\it et al.} (Particle Data Group) 
{\it Phys. Rev.} {\bf 86} (2012) 010001.

\bibitem{weinberg5}
S. Weinberg, {\it Phys. Rev.} {\bf 112} (1958) 1375.

\bibitem{bkm1994}
V. Bernard, N. Kaiser and U.-G. Meissner, 
{\it Phys. Rev. D} {\bf 50} (1994) 6899. 
 
\bibitem{fearing1997} 
H. W. Fearing, R. Lewis, N. Mobed and S. Scherer, 
{\it Phys. Rev. D} {\bf 56} (1997) 1783. 

\bibitem{ad66}
S. L. Adler and Y. Dothan, {\it Phys. Rev.} {\bf 151} (1966) 1267. 

\bibitem{wolfenstein}
L. Wolfenstein, in {\it High-Energy Physics and Nuclear Structure} 
ed. S. Devons 
(Plenum, New York, 1970) p. 661.

\bibitem{kaiser03} 
N. Kaiser, {\it Phys. Rev. C} {\bf 67} (2003) 027002.

\bibitem{raexp1999}
A. Liesenfeld {\it et al.} {\it Phys. Lett. B} {\bf 468} (1999) 20. 

\bibitem{czarnecki2007}
A. Czarnecki, W. J. Marciano and A. Sirlin, 
{\it Phys. Rev. Lett.} {\bf 99} (2007) 032003.

\bibitem{ando2000} 
S.-I. Ando, F. Myhrer and K. Kubodera, 
{\it Phys. Rev. C} {\bf 63} (2000) 015203. 

\bibitem{bhm2001} 
V. Bernard, T. R. Hemmert and U.-G. Meissner, 
{\it Nucl. Phys. A} {\bf 686} (2001) 290.

\bibitem{opat}
G.I. Opat, {\it Phys. Rev.} {\bf 134} (1964) B428. 

\bibitem{udit2013} 
U. Raha, F. Myhrer and K. Kubodera, 
{\it Phys. Rev. C} {\bf 87} (2013) 055501. 

\bibitem{EirasSoto}
D. Eiras and J. Soto, {\it Phys. Lett. B}, {\bf 491} (2000) 101.

\bibitem{Friar}
J.L. Friar, {\it Ann. Phys. (NY)}, {\bf 122} (1979) 151.

\bibitem{sirlin1967}
A. Sirlin, {\it Phys. Rev.} {\bf 164} (1967) 1767. 

\bibitem{MarSir86}
W.J. Marciano and A. Sirlin, {\it Phys. Rev. Lett.}, {\bf 56} (1986) 22.

\bibitem{ando2004}
S. Ando, H. W. Fearing, V. Gudkov, K. Kubodera, F. Myhrer, 
S. Nakamura and T. Sato, {\it Phys. Lett. B} {\bf 595} (2004) 250. 

\bibitem{Mund2013} 
D. Mund {\it et al.}, {\it Phys. Rev. Lett.} {\bf 110}  (2013) 172502.

\bibitem{Mendenhall2013}
M.P. Mendenhall {\it et al.} (UCNA Collaboration), 
{\it Phys. Rev. C} {\bf 87}  (2013) 032501.

\bibitem{greene2011} 
F.E. Wietfeldt and G.L. Greene, {\it Rev. Mod. Phys.} {\bf 83} (2011) 1173. 

\bibitem{ivanov2013} 
A.N. Ivanov {\it et al.}, arXiv:1306.1995 (2013). 

\bibitem{yue}
A.T. Yue {\it et al.}, {\it Phys. Rev. Lett.} {\bf 111} (2013) 222501.

\bibitem{Ericson2002}
T.E.O. Ericson, B. Loiseau and A.W. Thomas, 
{\it Phys. Rev. C} {\bf 66} (2002) 014005. 

\bibitem{Arndt2004}
R.A. Arndt, W.J. Briscoe, I.I. Strakovsky, R.L. Workman and M.M. Pavan, 
{\it Phys. Rev. C} {\bf 69} (2004) 035213 [arXiv:nucl-th/0311089]. 

\bibitem{Bugg2003} 
D.V. Bugg, {\it Eur. Phys. J. C} {\bf 33}  (2004) 505. 

\bibitem{Arndt2006}
R.A. Arndt, W.J. Briscoe, I.I. Strakovsky and R.L. Workman, 
{\it Phys. Rev. C} {\bf 74} (2006) 045205 [arXiv:nucl-th/0605082]. 
 
\bibitem{Baru2011}
V. Baru, C. Hanhart, M. Hoferichter, B. Kubis, A. Nogga and D. R. Phillips, 
{\it Phys. Lett. B} {\bf 694} (2011) 473.  

\bibitem{Baru2}
V. Baru, C. Hanhart, M. Hoferichter, B. Kubis, A. Nogga and D. R. Phillips,
{\it Nucl. Phys. A} {\bf 872} (2011) 69. 

\bibitem{Nijmegen1997}
J.J. de Swart, M.C.M. Rentmeester and R.G.E Timmermans,  
{\it PiN Newslett.} {\bf 13} (1997)  96 [arXiv:nucl-th/9802084]. 

\bibitem{KH83}
G. H\"{o}hler, in {\it Landolt -- B\"{o}rnstein} {\bf 9, b2} ed. H. Schopper 
(Springer, Berlin 1983).

\bibitem{saori}
S. Pastore, F. Myhrer and K. Kubodera, {\it Phys. Rev. C}
{\bf 88} (2013) 058501.



\bibitem{Ando2000}
  S.~Ando, T.~S.~Park, K.~Kubodera and F.~Myhrer,
  Phys.\ Lett.\ B {\bf 533}, 25 (2002)
  [nucl-th/0109053].


\bibitem{pisaPRL} 
  L.~E.~Marcucci, A.~Kievsky, S.~Rosati, R.~Schiavilla and M.~Viviani,
  Phys.\ Rev.\ Lett.\  {\bf 108}  (2012) 052502
  [arXiv:1109.5563 [nucl-th]].

\bibitem{pisa2011}
  L.~E.~Marcucci, M.~Piarulli, M.~Viviani, L.~Girlanda, A.~Kievsky, S.~Rosati, R.~Schiavilla,
  Phys.\ Rev.\ C\  {\bf 83}  (2011) 014002. 

\bibitem{Adam}
  J.~Adam, Jr., M.~Tater, E.~Truhlik, E.~Epelbaum, R.~Machleidt and P.~Ricci,
  Phys.\ Lett.\ B {\bf 709} (2012) 93
  [arXiv:1110.3183 [nucl-th]].

\bibitem{weinberg} 
S. Weinberg, {\it Physica A} {\bf 96} (1979) 327. 
\bibitem{weinberg2} 
S. Weinberg, {\it Phys. Lett. B} {\bf 251} (1990) 288. 
\bibitem{weinberg3} 
S. Weinberg, {\it Nucl. Phys. B} {\bf 363} (1991) 3. 
\bibitem{weinberg4} 
S. Weinberg, {\it Phys. Lett. B}  {\bf 295} (1992) 114. 

\bibitem{ParkMinRhoEM} 
T.-S. Park, D.-P. Min, M. Rho, Nucl.Phys. A596 (1996) 515-552.

\bibitem{ParkMinRhoAX} 
T.-S. Park, D.-P. Min, M. Rho, Phys.Rept. 233 (1993) 341-395.
%
%
\bibitem{Kolling09}
S.\ K\"olling, E.\ Epelbaum, H.\ Krebs, U.-G.\ Meissner,
Phys.\ Rev.\ {\bf C80}, 045502 (2009).
%
\bibitem{Kolling11}
S.\ K\"olling, E.\ Epelbaum, H.\ Krebs, and U.-G.\ Meissner,
Phys.\ Rev.\ C {\bf 84}, 054008 (2011).
%
\bibitem{Pastore08}
S.\ Pastore, R.\ Schiavilla, and J.L.\ Goity,
Phys.\ Rev.\ C {\bf 78}, 064002 (2008).
%
\bibitem{Pastore09}
S. Pastore, L. Girlanda, R. Schiavilla, M. Viviani, and R. B. Wiringa,
Phys. Rev. C {\bf 80}, 034004 (2009).
%
%
\bibitem{Piarulli12}
M.\ Piarulli, L.\ Girlanda, L.E.\ Marcucci, S.\ Pastore, R.\ Schiavilla, and M. \ Viviani,
Phys. Rev. C {\bf 87}, 014006 (2013).

\bibitem{Baroni14}
A. Baroni {\it et al.},     
in preparation.

\bibitem{park2}
 T.~S.~Park {\it et al.},
  Phys.\ Rev.\  C {\bf 67} (2003) 055206
  [arXiv:nucl-th/0208055].

\bibitem{marcucci2013}
L.E. Marcucci, R. Schiavilla and M. Viviani, 
{\it Phys. Rev. Lett.} {\bf 110} (2013) 192503. 

\bibitem{Nakamura:2002jg} 
  S.~Nakamura, T.~Sato, S.~Ando, T.~S.~Park, F.~Myhrer, V.~P.~Gudkov and K.~Kubodera,
  Nucl.\ Phys.\ A {\bf 707} (2002) 561 
  [nucl-th/0201062].


\bibitem{HvKM2000}
C. Hanhart, U. van Kolck and G.A. Miller, 
{\it Phys. Rev. Lett.} {\bf 85} (2000) 2905. 

\bibitem{review}
V. Baru, C. Hanhart and F. Myhrer, 
 {\it Int. J. Mod. Phys.} {\bf 23} (2014) 1430004    
[arXiv:1310.3505]. 

\bibitem{gazit}
  D.~Gazit, S.~Quaglioni and P.~Navratil,
  Phys.\ Rev.\ Lett.\  {\bf 103} (2009) 102502
  [arXiv:0812.4444 [nucl-th]].

\bibitem{pdchiral}
  E.~Epelbaum, A.~Nogga, W.~Gloeckle, H.~Kamada, U.-G.~Mei\ss ner and H.~Witala,
  Phys.\ Rev.\  C {\bf 66} (2002) 064001
  [arXiv:nucl-th/0208023].

\bibitem{Epelbaum:2008ga}
  E.~Epelbaum, H.~-W.~Hammer and U.~-G.~Meissner,
  Rev.\ Mod.\ Phys.\  {\bf 81} (2009) 1773
  [arXiv:0811.1338 [nucl-th]].
  
  \bibitem{Machleidt:2011zz}
  R.~Machleidt and D.~R.~Entem,
  Phys.\ Rept.\  {\bf 503} (2011) 1
  [arXiv:1105.2919 [nucl-th]].

\bibitem{gammad}
  V.~Lensky, V.~Baru, J.~Haidenbauer, C.~Hanhart, A.~E.~Kudryavtsev and U.~G.~Meissner,
  Eur.\ Phys.\ J.\  A {\bf 26}, 107 (2005)
  [arXiv:nucl-th/0505039]

\bibitem{gammad_nn}
  V.~Lensky, V.~Baru, E.~Epelbaum, C.~Hanhart, J.~Haidenbauer, 
A.~E.~Kudryavtsev and U.~G.~Meissner,
  Eur.\ Phys.\ J.\  A {\bf 33}, 339 (2007)
  [arXiv:0704.0443 [nucl-th]].

\bibitem{gardestig}
 A.~Gardestig,
  Phys.\ Rev.\  C {\bf 74} (2006) 017001
  [arXiv:nucl-th/0604035].

\bibitem{phillips}
 A.~Gardestig and D.~R.~Phillips,
  Phys.\ Rev.\ Lett.\  {\bf 96} (2006) 232301.
  [arXiv:nucl-th/0603045].

\bibitem{Gardestig:2005pp}
  A.~Gardestig and D.~R.~Phillips,
  Phys.\ Rev.\ C {\bf 73} (2006) 014002
  [nucl-th/0501049].

\bibitem{gardestigreview}
  A.~Gardestig,
  J.\ Phys.\ G {\bf 36} (2009) 053001
  [arXiv:0904.2787 [nucl-th]].

\bibitem{Argonne}
R.B. Wiringa, V.G.J. Stoks and R. Schiavilla,
{\it Phys. Rev. C} {\bf 51} (1995) 38.

\bibitem{3N}
B.S. Pudliner {\it et al.}, {\it Phys. Rev. C} {\bf 56} (1997) 1720. 

\bibitem{Rocco} 
J. Carlson and R. Schiavilla, {\it Rev. Mod. Phys.} {\bf 70} (1998) 743. 

\bibitem{Tatara}
N. Tatara, Y. Kohyama and K. Kubodera, {\it Phys. Rev. C} {\bf 42} (1990) 1694.  

\bibitem{Adam-Q}
J. Adam, E. Truhlik, S. Ciechanowicz and K.-M. Schmitt, {\it Nucl. Phys. A} {\bf 507} (1990) 675. 


\bibitem{gardestig07} 
A. G\aa rdestig, K. Kubodera and F. Myhrer, 
{\it Phys. Rev. C} {\bf 76} (2007) 014005. 

\bibitem{Truhlik2010}
P. Ricci, E. Truhlik, B. Mosconi, J. Smejkal, {\it Nucl. Phys. A} {\bf 837} (2010) 110. 


\bibitem{Song2013}
Y.-H. Song     
{\it et al.},  
in preparation.

\bibitem{gazit1}
D.~Gazit, {\it Phys. Lett. B} {\bf 666} (2008) 472.

\bibitem{RMCexp}
D. H. Wright {\it et al.} {\it Phys. Rev. C} {\bf 57} (1998) 373.

\bibitem{meissner1998}
T. Meissner, F. Myhrer and K. Kubodera, {\it Phys. Lett. B} {\bf 416} (1998) 36.

\bibitem{ando1998}
S. Ando and D.-P. Min, {\it Phys. Lett. B} {\bf 417} (1998) 177. 













\end{thebibliography}
\end{document}